\documentclass[preprint,showpacs,preprintnumbers,amsmath,amssymb]{revtex4}

\usepackage{graphicx}
\usepackage{epsfig}		
\usepackage{dcolumn}
\usepackage{bm}

\def\0{\mbox{\tiny $0$}}
\def\1{\mbox{\tiny $1$}}
\def\2{\mbox{\tiny $2$}}
\def\3{\mbox{\tiny $3$}}
\def\4{\mbox{\tiny $4$}}
\def\5{\mbox{\tiny $5$}}
\def\6{\mbox{\tiny $6$}}
\def\7{\mbox{\tiny $7$}}
\def\8{\mbox{\tiny $8$}}
\def\9{\mbox{\tiny $9$}}
\def\R{\mbox{\tiny $R$}}

\def\bb#1{\mbox{\footnotesize $(#1)$}}
\def\x{\mbox{\tiny $x$}}

\def\L{\mbox{\tiny $L$}}

\def\D{\mbox{\tiny $D$}}

\def\pmp{\mbox{\tiny $\mp$}}

\begin{document}
\title{Relation between phase and dwell times for quantum tunneling of a relativistically propagating particle}

\author{A. E. Bernardini}
\email{alexeb@ifi.unicamp.br}
\affiliation{Instituto de F\'{\i}sica Gleb Wataghin, UNICAMP,
PO Box 6165, 13083-970, Campinas, SP, Brasil.}

\date{\today}

\begin{abstract}
The general and explicit relation between the phase time and the dwell time for quantum tunneling of a relativistically propagating particle is investigated and quantified.
In analogy with previously obtained non-relativistic results, it is shown that the group delay can be described in terms of the dwell time and a self-interference delay.
Lessons concerning the phenomenology of the relativistic tunneling are drawn.
\end{abstract}

\pacs{03.65.Xp - 03.65.Pm}
\keywords{Phase Time - Dwell Time - Tunnel Effect - Relativistic Wave Equation}
\date{\today}
\maketitle

To obtain the definitive answer for the time spent by particle to penetrate a classically forbidden region delimited by a potential barrier \cite{Con70,But83,Hau87,Fer90}, people have tried to introduce quantities that have the dimension of time and can somehow be associated with the passage of the particle through the barrier or, strictly speaking, with the definition of the tunneling times \cite{Bro94,Sok94,Jak98,Olk02,WinWin,Winful,Olk04,Ber08}.
Tunneling is a general feature of wave equations and may be very counterintuitive when compared with the evolution of propagating waves.
It occurs when a wave impinges on a thin barrier of opaque material and some small amount of the wave {\em leaks} through to the other side.
Even so, it has been mainly discussed for the Schroedinger equation due to the shocking contrast between classical and quantum particles.
In all cases described by the non-relativistic (Schroedinger) dynamics \cite{Olk04}, the pulse (wave packet) that emerges from the tunneling process is greatly attenuated and front-loaded due to the {\em filter} effect (only the leading edge of the incident wave packet survives the tunneling process without being severally attenuated to the point that it cannot be detected).
If one measures the speed by the peak of the pulse, it looks faster than the incident wave packet.
Moreover, since the transmission probability depends analytically on the momentum component $k$  ($T \equiv T\bb{k}$), the initial (incident wave) momentum distribution can be completely distorted by the presence of the barrier of potential.

In what concerns the momentum distribution distortion and the precise computation of phase times, considering the relativistic tunneling dynamics in terms of the Dirac/Klein-Gordon wave equation allows for circumventing such difficulties.
Even though in the non-relativistic framework \cite{Winful}, a quite elegant study is performed so to overcome the above mentioned misunderstanding of the tunneling time definitions.
Indeed some authors consider difficult and perhaps confusing the treatment of all interactions of plane waves or wave packets with a barrier potential using a relativistic wave equation \cite{Del03,Cal99,Dom99,Che02}.
This is because the physical content depends upon the relation between the barrier height $V_{\0}$ and the mass $m$ of the incoming (particle) wave, beside of its total energy $E$.
In some previous analysis \cite{Ber07A}, we have demonstrated with complete mathematical accuracy that, in some limiting cases of the relativistic (Klein-Gordon) tunneling phenomena where the relativistic kinetic energy is approximately equal to the potential energy of the barrier, and $m c L /\hbar << 1$, particles with mass $m$ can pass through a potential barrier $V_{\0}$ of width $L$ with transmission probability $T$ approximately equal to the unity (total transmission).

Differently from other previous (non-relativistic) tunneling analysis, the original momentum distribution is kept undistorted and there is no {\em filter} effect.
The tunneling time is then computed for a completely undistorted transmitted wave packet, which legitimizes any eventual accelerated transmission \cite{Ber07A}.

Turning back to the first attempt of evaluating this problem, Klein \cite{Kle29} considered the reflection and the transmission of electrons of energy $E$ incident on the potential step $V\bb{x} = \Theta\bb{x}V_{\0}$ in the one-dimensional time-independent Dirac equation which can be represented in terms of the usual Pauli matrices \cite{Zub80} by\footnote{$\Theta\bb{x}$ is the Heavyside function.}
\small\begin{equation}
\left[\sigma^{3}\sigma^{\x}\partial_{\x} - (E - \Theta\bb{x}V_{\0}) - \sigma^{z} m\right]\phi\bb{k, x} = 0,
~~(\mbox{from this point}~ c = \hbar = 1).
\label{001}
\end{equation}\normalsize
which corresponds to the reduced representation of the usual Pauli-Dirac {\em gamma} matrix representation obtained when the spinorial character is neglected ($1+1$ dimensional Dirac equation).
The physical essence of such a theoretical configuration lies in the prediction that fermions can pass through large repulsive potentials without exponential damping, in a kind of (Klein) tunnelling phenomenon \cite{Cal99} which follows accompanied by the production of a particle-antiparticle pair inside the potential barrier.
It is different from the usual tunneling effect since it occurs inside the energy zone of the Klein paradox \cite{Kle29,Zub80}.

Taking the quadratic form of the $1+1$ dimensional Dirac equation, we obtain the Klein-Gordon equation for the time-like component $V\bb{x}$ of a Lorentz four-vector potential,
\small\begin{equation}
\left(E - V\bb{x}\right)^{\2}\phi\bb{k, x} = \left(-\partial^{\2}_{\x} + m^{\2}\right)\phi\bb{k, x},
\label{002}
\end{equation}\normalsize
which, from the mathematical point of view, due to the second-order spatial derivatives, has boundary conditions similar to those ones of the Schroedinger equation and leads to stationary wave solutions characterized by a {\em relativistically} modified dispersion relation.

All these proposals for computing how long a particle takes to tunnel through a potential barrier have led to the introduction of several {\em transit time} definitions, among which, in spite of no general agreement \cite{Olk04}, the so called phase time \cite{Wig55} (group delay) and the dwell time have an apparently well established quantified relation \cite{Hau89,Win03} for non-relativistic Schroedinger equation solutions.
In this manuscript, we extend such results to the one-dimensional scattering potential configuration described by Klein-Gordon equation solutions.

Let us then depict the three potential regions by means of a rectangular potential barrier $V\bb{x}$, $V\bb{x} = V_{\0}$ if $0 \leq x \leq L$, and $V\bb{x} = 0$ if $x < 0$ and $x > L$.
Differently from the non-relativistic (Schroedinger) dynamics, we observe that the incident energy can be divided into three zones.
The {\em above barrier} energy zone, $E > V_{\0} + m$, involves diffusion phenomena of oscillatory waves (particles).
In the so called {\em Klein} zone \cite{Kle29,Cal99}, $E < V_{\0} - m$, we find oscillatory solutions (particles and antiparticles) in the barrier region.
In this case, antiparticles see an electrostatic potential opposite to that seen by the particles and hence they will see a well potential where the particles see a barrier \cite{Aux1,Kre04}.
The {\em tunnelling} zone, $V_{\0} - m < E < V_{\0} + m$, for which only evanescent waves exist \cite{Kre01,Pet03} in the barrier region, is that of interest in this work.
By evaluating the problem for this tunneling (evanescent) zone assuming that $\phi (k,x)$ are stationary wave solutions of Eq.~(\ref{002}), when the peak of an incident (positive energy) wave packet reach the barrier $x = 0$ at $t = 0$, we can usually write
\small\begin{equation}
\phi(k,x)=
\left\{\begin{array}{l l l l}
\phi_{\1}(k,x) &=&
\exp{\left[ i \,k \,x\right]} + R(k,L)\exp{\left[ - i \,k \,x \right]}&~~~~x < 0,\nonumber\\
\phi_{\2}(k,x) &=& \alpha(k)\exp{\left[ - \rho\bb{k}  \,x\right]} + \beta(k)\exp{\left[ \rho\bb{k}  \,x\right]}&~~~~0 < x < L,\nonumber\\
\phi_{\3}(k,x) &=& T(k,L)\exp{ \left[i \,k (x - L)\right]}&~~~~x > L,
\end{array}\right.
\label{003}
\end{equation}\normalsize
where the dispersion relations are modified with respect to the usual non-relativistic ones: $k^{\2} = E^{\2} - m^{\2}$ and $ \rho\bb{k}^{\2} = m^{\2} - (E - V_{\0})^{\2}$.

To establish a correspondence with the non-relativistic (NR) solutions, it is convenient to define the kinematic variables in terms of the following parameters: $w = \sqrt{2 m V_{\0}}$, $\upsilon = V_{\0}/m = w^{\2}/2m^{\2}$, and $n^{\2}\bb{k} = k^{\2}/w^{\2} = E_{NR}/V_{\0}$.
The parameter $w$ corresponds to the same {\em normalizing} parameter of the usual NR analysis where $k^{\2} = 2 m E_{NR}$.
The above mentioned relation between the potential energy $V_{\0}$ and the mass $m$ of the incident particle is given by the parameter $\upsilon$.
Finally, $n^{\2}\bb{k}$ represents the dependence on the energy for all the results that will be considered here.
After simple mathematical manipulations, it is easy to demonstrate that the tunneling zone for the above form of the Klein-Gordon equation (\ref{002}) is comprised by the interval $(n^{\2}\bb{k} - \upsilon/2)^{\2} \leq 1$ for which $n^{\2}\bb{k}$ might assume larger values ($n^{\2}\bb{k} >> 1$), in opposition to the NR case where the tunneling energy zone is constrained by $0 < n^{\2}\bb{k} < 1$).
We shall observe that such a peculiarity has a subtle relation with the possibility of superluminal transmission through the barrier.
The limits for NR energies ($k^{\2} << m^{\2}$ and $V << m$) are given by $\upsilon n << 1$ and $\upsilon/n << 1$, which, as we have indicated in a previous analysis \cite{Ber07A}, reproduces the transmission and delay results of the Schroedinger equation.

The stationary phase method can be successfully applied for describing the movement of the center of a wave packet constructed in terms of a symmetrical momentum distribution $g(k - k_{\0})$ which has a pronounced peak around $k_{\0}$.
By assuming that the phase that characterizes the propagation varies smoothly around the maximum of $g(k - k_{\0})$, the stationary phase condition enables us to calculate the position of the peak of the wave packet (highest probability region to find the propagating particle).
With regard to the {\em standard} one-way direction wave packet tunneling, for the set of stationary wave solutions given by Eq.~(\ref{003}), it is well-known \cite{Ber06} that the transmitted amplitude $T\bb{n, L} = |T\bb{n, L}|\exp{[i \varphi\bb{n, L}]}$ is written in terms of
\small\begin{equation}
|T\bb{n, L}| = \left\{1 + \frac{1}{4 \, n^{\2} \, \rho^{\2}\bb{n}} \sinh^{\2}{\left[\rho\bb{n}\, w L \right]}\right\}^{-\frac{1}{2}},
\label{004}
\end{equation}\normalsize
where we have suppressed from the notation the dependence on $k$, and
\small\begin{equation}
\varphi\bb{n, L} = \arctan{\left\{\frac{n^{\2} - \rho^{\2}\bb{n}}
{2 n \, \rho\bb{n}}
\tanh{\left[\rho\bb{n} \, w L \right]}\right\}},
\label{005}
\end{equation}\normalsize
for which we have made explicit the dependence on the barrier length $L$ (parameter $w L$) and we have rewritten $\rho\bb{k} = w \,\rho\bb{n}$, with $\rho\bb{n}^{\2} = \sqrt{1 + 2 n^{\2} \upsilon} - (n^{\2} -\upsilon/2)$.

The additional phase $\varphi(n, L)$ that goes with the transmitted wave is utilized for calculating the transit time $t^{(\varphi)}$ of a transmitted wave packet when its peak emerges at $x = L$,
\small\begin{equation}
t_{\varphi} = \frac{\mbox{d}k}{\mbox{d}E\bb{k}} \frac{\mbox{d}n\bb{k}}{\mbox{d}k} \frac{\mbox{d}\varphi\bb{n, L}}{\mbox{d}n} = \frac{L}{v} \frac{1}{w\,L} \frac{\mbox{d}\varphi\bb{n, L}}{\mbox{d}n},
\label{006}
\end{equation}\normalsize
evaluated at $k = k_{\0}$ (the maximum of a generic symmetrical momentum distribution $g(k - k_{\0})$ that composes the {\em incident} wave packet).
By introducing the {\em classical} traversal time defined as
$\tau_{\bb{k}} = L (\mbox{d}k/\mbox{d}E\bb{k})= L / v$,
we can obtain the normalized phase time,
\small\begin{equation}
\frac{t_{\varphi}}{\tau_{\bb{k}}}
 = \frac{f\bb{n, L}}{g\bb{n, L}},
\label{007}
\end{equation}\normalsize
where
\small\begin{eqnarray}
f\bb{n, L} &=&
8 n^{\2} \left[\left(2 + 8 n^{\2} \upsilon + \upsilon^{\2}\right) - \left(4 n^{\2} + 3 \upsilon\right)\sqrt{1 + 2 n^{\2}\upsilon} \right]\nonumber\\
&&~~~~+
4 \left[\left(4 + 4 n^{\2}\upsilon + \upsilon^{\2}\right)\sqrt{1 + 2 n^{\2}\upsilon} - 2 \upsilon \left(2 + 3 n^{\2}\upsilon\right)\right]\frac{Sh(\rho\bb{n} w L)\,Ch(\rho\bb{n} w L)}{\rho\bb{n} w L},
\nonumber\\
g\bb{n, L} &=&
16 n^{\2} \left[2 \left(1 + 2 n^{\2}\upsilon\right) -  \sqrt{1 + 2 n^{\2}\upsilon}\left(2 n^{\2} + \upsilon\right)\right]\nonumber\\
&&~~~~+
2 \left[\left(4 + 8 n^{\2}\upsilon + \upsilon^{\2}\right)\sqrt{1 + 2 n^{\2}\upsilon} - 4 \upsilon \left(1 + 2 n^{\2}\upsilon\right)\right]Sh(\rho\bb{n} w L)^{\2},
\nonumber
\end{eqnarray}\normalsize
with $Ch(x) = \cosh{(x)}$ and $Sh(x) = \sinh{(x)}$.

Turning back to the main point, could one say metaphorically that the particle represented by the positive energy incident wave packet spend a time equal to $ t_{T, \varphi}$ inside the barrier before retracing its steps or tunneling?
The answer is in the definition of the dwell time for the relativistic colliding configuration which we have proposed.
In quantum mechanics, using steady-state wave functions, the average time of residence in a region is the integrated density divided by the total flux in (or out) and the lifetime is defined as the difference between these residence times with and without interactions.
In non-relativistic quantum mechanics, the dwell time is a measure of the time spent by a particle in the barrier region regardless of whether it is ultimately transmitted or reflected \cite{But83},
\small\begin{equation}
t^{(\D)}
=\frac{1}{j_{in}} \int_{\0}^{\L}{\mbox{d}x{|\phi_{\2}(k,x)|^{\2}}},
\label{008}
\end{equation}\normalsize
where  $j_{in} = k / m$ is the flux of positive energy incident particles and $\phi_{\2}(k,x)$ is the stationary state wave function inside the barrier.

In terms of the redefined parameters $n$ and $\upsilon$, the explicit expression for the dwell time normalized by $\tau_{(k)}$ is given by
\small\begin{equation}
\frac{t^{(\D)}}{\tau_{(k)}} = \frac{f_{\D}\bb{n, L}}{g_{\D}\bb{n, L}}
\label{009}
\end{equation}\normalsize
where
\small\begin{eqnarray}
f_{\D}\bb{n, L} &=&\left(1 - \frac{n^{\2}}{\rho\bb{n}^{\2}}\right) + \left(1 + \frac{n^{\2}}{\rho\bb{n}^{\2}}\right) \frac{Sh(\rho\bb{n} w L)\,Ch(\rho\bb{n} w L)}{\rho\bb{n} w L}
\nonumber\\
g_{\D}\bb{n, L} &=& 2\sqrt{1 + 2 n^{\2}\upsilon} \left[1 + \frac{Sh(\rho\bb{n} w L)^{\2}}{4 n^{\2} \rho\bb{n}^{\2}}\right]
\nonumber
\end{eqnarray}\normalsize

To derive the relation between the dwell time and the phase time, we reproduce the variational theorem which yields the sensitivity of the wave function to variations in energy.
Following from the Smith derivation \cite{Smi60} for the non-relativistic Schroedinger equation, here we also have the eigenvalue equation
\small\begin{equation}
\left(i \partial_{\0} - E \right)\phi\bb{k, x} = 0,
\label{002A}
\end{equation}\normalsize
and its first derivative with respect to $E$,
\small\begin{equation}
\left(i \partial_{\0} - E \right)\frac{\partial \phi\bb{k, x}}{\partial E} - \phi\bb{k, x } = 0.
\label{002B}
\end{equation}\normalsize
After some simple mathematical manipulations and the substitution of the Klein-Gordon equation (\ref{002}), the second derivative can be written as
\small\begin{equation}
\left(\partial^{\2}_{\0} + E^{\2} \right)\frac{\partial \phi\bb{k, x}}{\partial E} + 2 E \phi\bb{k, x} = 0.
\label{002C}
\end{equation}\normalsize
By following the one-dimensional analysis here considered, it is easy to find that
\small\begin{eqnarray}
\left[\frac{\partial \phi}{\partial E} \frac{\partial^{\2}}{\partial x^{\2}}\phi^{\dagger} - \phi^{\dagger}\frac{\partial^{\2}}{\partial x^{\2}}\frac{\partial \phi}{\partial E}\right] &=& 2(E - V_{\0}) \phi^{\dagger}\phi
=\frac{\partial}{\partial x}\left(\frac{\partial \phi}{\partial E}\frac{\partial \phi^{\dagger}}{\partial x} - \phi^{\dagger}\frac{\partial^{\2}\phi}{\partial E\partial x}\right),
\label{020}
\end{eqnarray}\normalsize
where we clearly notice the presence of $E - V_{\0}$ in place of $m$ of the non-relativistic result (\ref{008}) \cite{Smi60}.
Upon integration over the barrier length we find
\small\begin{equation}
2 (E - V_{\0}) \int_{\0}^{{\L}}\mbox{d}x{|\phi_{\2}(k,x)|^{\2}} = \left.\left(\frac{\partial \phi}{\partial E}\frac{\partial \phi^{\dagger}}{\partial x} - \phi^{\dagger}\frac{\partial^{\2}\phi}{\partial E\partial x}\right)\right|_{\0}^{\L}.
\label{021}
\end{equation}\normalsize
In front of the barrier ($x \leq 0$), the wave function consists of an incident and a reflected component given by $\phi_{\1}\bb{k, x}$, and
behind the barrier ($x \leq L$), there is only the transmitted wave $\phi_{\3}\bb{k, x}$ (see Eq.~(\ref{003})).
Under these conditions we evaluate the right-hand side of Eq.~(\ref{021}) as
\small\begin{equation}
-2 i k \left[\frac{\mbox{d}}{\mbox{d}k}(|R|^{\2}+|T|^{\2}) + i \left((|R|^{\2}+|T|^{\2})\frac{\mbox{d}\varphi\bb{k, L}}{\mbox{d}k} + \frac{Im[R]}{k}\right)\right]\frac{\mbox{d}k}{\mbox{d}E}.
\label{022}
\end{equation}\normalsize
Since $|R|^{\2}+|T|^{\2} = 1$, Eq.~(\ref{021}) becomes
\small\begin{equation}
(E - V_{\0}) \int_{0}^{{\L}}\mbox{d}x{|\phi_{\2}(k,x)|^{\2}} =
\frac{\mbox{d}\varphi\bb{k, L}}{\mbox{d}E} + \frac{k}{E} Im[R]
\label{023}
\end{equation}\normalsize
which gives
\small\begin{equation}
\frac{t^{(\varphi)}}{\tau_{(k)}} = \frac{t^{(\D)}_{\R}}{\tau_{(k)}} - \frac{1}{\tau_{(k)}}\frac{Im[R]}{E}
\label{024}
\end{equation}\normalsize
where we have introduced the {\em re-scaled} dwell time,
\small\begin{equation}
t^{(\D)}_{\R}  = \frac{E - V_{\0}}{m} t^{(\D)} = \frac{E - V_{\0}}{k} \int_{\0}^{\L}{\mbox{d}x{|\phi_{\2}(k,x)|^{\2}}}
\label{025}
\end{equation}\normalsize
which can be related to the correct definition of the probability density for the Klein-Gordon equation,
\small\begin{equation}
j_{\0} = \int_{\0}^{\L}{\mbox{d}x\,[\phi^{\dagger}_{\2}(\partial_{\0}\phi_{\2}) - (\partial_{\0}\phi^{\dagger}_{\2})\phi_{\2}}]/ ~~~~ t^{(\D)}_{\R} = (j_{\0}/j_{in}) = (j_{\0}/ k),
\label{025B}
\end{equation}\normalsize
and leads to the usual definition $t^{(\D)}$ in the non-relativistic limit $(\frac{E - V_{\0}}{m} \mapsto 1)$.
In Eq.~(\ref{025}), the squared modulus of the wave function transforms as a Lorentz scalar,
$E - V_{\0}$ transforms as a time-like component, and the integrand $dx$ as well as $k$ transform as space-like components of a Lorentz four-vector.
It means that $t^{(\D)}_{\R}$ has the correct Lorentz character since it transforms as a time-like component, which does not occur for $t^{(\D)}$ of Eq.~(\ref{008}).

\begin{figure}[th]
\vspace{-0.3 cm}
\centerline{\psfig{file=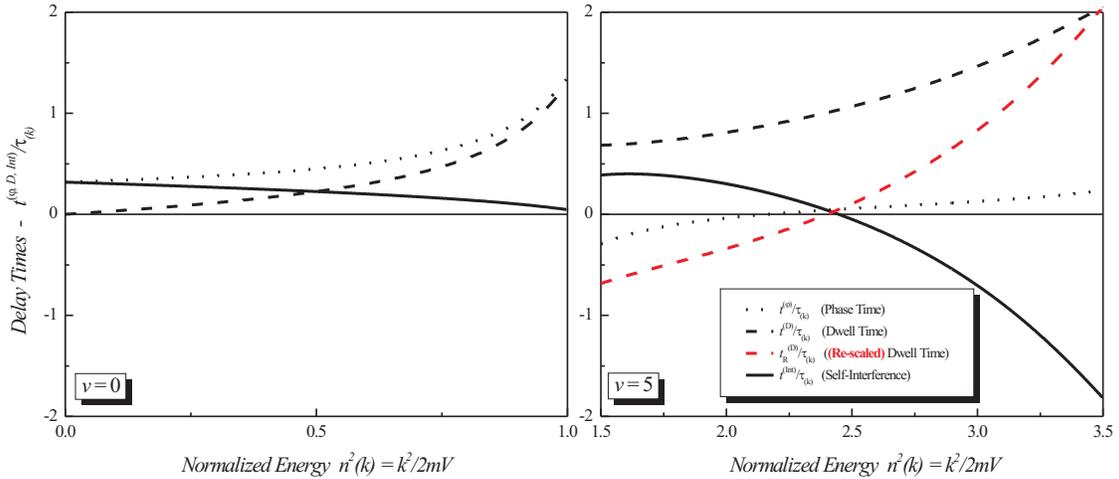,width=17cm}}
\vspace{-0.5 cm}
\caption{Delay times calculated from the dynamics of the Klein-Gordon equation: Phase Time (Dash-dotted line), Dwell Time (Dashed black line), Self-Interference term (Solid line), and the {\em re-scaled} Dwell Time (Dashed red line).
In fact the tunneling region is comprised by the interval $(n^{\2} - \upsilon/2)^{\2} < 1,\, n^{\2} > 0$.
Here we have adopted the illustratively convenient value of $w L = 2 \pi$ with $\upsilon =  5$, in comparison with the non-relativistic results parameterized by $\upsilon \rightarrow 0$.
\label{Fig02}.}
\end{figure}

As in the non-relativistic case \cite{Winful, Smi60}, the first term of Eq.~(\ref{024}) corresponds to the phase time or the aforementioned group delay.
The second term comes from the explicit computation of the dwell time.
However, the presence of the multiplicative factor $\frac{E - V_{\0}}{m}$ in Eq.~(\ref{025}) introduces some novel aspects in the interpretation of the additional term $-Im[R] / E$ as a self-interference term which comes from the momentary overlap of incident and reflected waves in front of the barrier \cite{WinWin}.
As we can observe in the example illustrated in the Fig.~\ref{Fig02}, and by the usual definition (\ref{008}), the dwell time is always positive.
The {\em re-scaled} dwell time modulated by $(\frac{E - V_{\0}}{m}$ changes sign when the total energy $E$ equalizes the potential energy $V_{\0}$: an energy region comprised by the tunneling energy zone of the Klein-Gordon equation.
Consequently, differently from the results we get from the non-relativistic analysis, the dwell time is not obtained from a simple subtraction of the {\em supposed} self-interference delay $t^{\bb{Int}}$ from the phase time that, in some circumstances \cite{Ber06,Ber07A} describes the exact position of the peak of the scattered wave packets.
Such results give a complete description of the the {\em tunneling} zone, $V_{\0} - m < E < V_{\0} + m$, for which only evanescent waves exist \cite{Kre01,Pet03}, several times ignored in the analysis of relativistic tunneling.
For the evanescent tunneling zone, the Dirac equation and its quadratic form (namely, the Klein-Gordon  equation) leads to the same results when we apply the (evanescent) tunneling time definitions in which we are interested, the phase time and the dwell time.
The evanescent tunneling zone does not intersect with the Klein paradox energy zone for which, at least theoretically, the possibility of creation/annihilation of fermionic pairs leads to the reinterpretation of the probability density currents, and thus to a novel interpretation of the tunneling phenomenon.
Consequently, concerning the calculation of evanescent tunneling times, all the references to fermionic (Dirac) and bosonic (Klein-Gordon) particles are valid, in the same sense that all the results derived from the non-relativistic Schr\"{o}dinger equation are supposed to be valid for massive fermions and massive bosons.

At least for the moment, the above (relativistic) results do not necessarily demand for a confront with the (non-relativistic) predictions derived from the opaque limit analysis which results in the filter effect and the superluminal tunneling.
To clear up this assertion, it is convenient to recover the limiting configurations ($n^{\2}\rightarrow \upsilon/2 \pmp 1$) of some of our previous results \cite{Ber07A} for which the tunneling transmission probability (\ref{004}) can be approximated by
\small\begin{equation}
\lim_{n^{\2}\rightarrow \upsilon/2 \pmp 1}{|T\bb{n, L}|} =
\left[1 + \frac{(w L)^{\2}}{2 \upsilon \mp 4}\right]^{-\frac{1}{2}}
\mbox{$\begin{array}{c}\mbox{\tiny$\upsilon >> 1$}\\ \rightarrow \\~\end{array}$}
\left[1 + (m L)^{\2}\right]^{-\frac{1}{2}},
\label{011}
\end{equation}\normalsize
from which, avoiding any kind of filter effect, we recover the probability of complete tunneling transmission when $m L << 1$, once we have $|T\bb{n, L}| \approx 1$.
For the correspondent values of the phase times we obtain \cite{Ber07A},
\small\begin{equation}
\lim_{n^{\2}\rightarrow \upsilon/2 \mp 1}{\frac{t^{(\varphi)}}{\tau_{(k)}}} = -\frac{4}{3}\frac{1}{1 \pm 2 n^{\2}}, ~~~~ n^{\2}\rightarrow \upsilon/2 \mp 1, ~~~n^{\2},\,\upsilon > 0,
\label{012}
\end{equation}\normalsize
that does not depend on $m L$, and we notice that its asymptotic (ultrarelativistic) limit always converges to $0$.
Curiously, in the lower limit of the tunneling energy zone, $n^{\2}\rightarrow \upsilon/2 - 1$, it is always negative.
Since the result of Eq.~(\ref{012}) is exact, and we have accurately introduced the possibility of obtaining total transmission ({\em transparent barrier}), our result ratifies the possibility of accelerated transmission (positive time values), and consequently superluminal tunneling (negative time values), for relativistic particles when $m L$ is sufficiently smaller than 1 $(\Rightarrow T \approx 1)$.
By observing that the barrier height has to be chosen such that one remains in the tunneling regime,
it is notorious that the transmission probability depends only weakly on the barrier height, approaching the perfect transparency for very high barriers, in stark contrast to the conventional, non-relativistic tunneling where $T\bb{n, L}$ exponentially  decays with the increasing $V_{\0}$.
Obviously, the above results correspond to a theoretical prediction, in certain sense, not so far from the experimental realization.
The above condition should be naturally expected since we are simply assuming that the Compton wavelength ($\hbar/(m c)$) is much larger than the length $L$ of the potential barrier that, in this case, becomes {\em invisible} for the tunneling particle.
In general terms, the relativistic quantum mechanics establishes that if a wave packet is spread out over a distance $d >> 1/m$, the contribution of momenta $|p| \sim m >> 1/d$ is heavily suppressed, and the negative energy components of the wave packet solution are negligible; the one-particle theory is then consistent.
If we want to localize the wave packet in a region of space (wave packet width $d$) smaller than or of the same size as the Compton wavelenght, that is $d < 1/m$, the negative energy solutions (antiparticle states) start to play an appreciable role.
The condition $d < L < 1/m$ (where $d < L$ is not mandatory) imposed over a positive energy component of the incident wave packet in the relativistic tunneling configuration excite the negative energy modes (antiparticles) and, qualitatively, report us to the Klein paradox and the creation of particle-antiparticles pairs during the scattering process which might create the intrinsic (polarization) mechanisms for accelerated and/or non-causal particle teletransportation.

To conclude, in analogy with previous non-relativistic results \cite{Winful}, we have shown that the group delay can be described in terms of the dwell time and a self-interference delay.
The general and explicit relation between phase times and the dwell times for quantum tunneling of a relativistically propagating particle was investigated and quantified.
Our analysis corroborates with the statement of conditions for the occurrence of accelerated tunneling transmission probabilities at nanoscopic scale in confront with the problematic superluminal interpretation originated from the study based on non-relativistic dynamics of tunneling \cite{Ber07A}.
By eliminating the filter effect, the transmission probabilities approximates the unitary modulus (complete transmission through a {\em transparent} medium).
In this case, we have noticed the possibility of accelerated ($t_{\varphi} < \tau_{\bb{k}}$), and eventually {\em superluminal} (negative tunneling delays, $t_{\varphi} < 0$) transmissions without recurring to the usual analysis of the {\em opaque} limit ($\rho\bb{n} w L \rightarrow \infty$) which leads to the Hartman effect \cite{Har62}.

{\bf Acknowledgments}
We would like to thank FAPESP (PD 04/13770-0) for the financial support.


\begin{thebibliography}{99}

\bibitem{Con70}
J. N. L. Connor, Molec. Phys. {\bf 19}, 65 (1970);
J. N. L. Connor, Molec. Phys. {\bf 15}, 37 (1968).
\bibitem{But83}
M. B\"{u}ttiker, Phys. Rev. {\bf B27}, 6178 (1983).
\bibitem{Hau87}
E. H. Hauge, J. P. Falck, and T. A. Fjeldly, Phys. Rev. {\bf B36}, 4203 (1987).
J. P. Falck and E.H. Hauge, Phys. Rev. {\bf B38}, 3287 (1988).
\bibitem{Fer90}
H. A. Fertig, Phys. Rev. Lett. {\bf 65}, 234 (1990).
\bibitem{Bro94}
S. Brouard, R. Sala, and J. G. Muga, Phys. Rev. {\bf A49}, 4312 (1994).
\bibitem{Sok94}
D. Sokolovski and J. N. L. Connor, Solid State Communications {\bf 89}, 475 (1994).
\bibitem{Jak98}
J. Jakiel, V. S. Olkhovsky, and E. Recami, Phys. Lett. {\bf A248}, 156 (1998).
\bibitem{Olk02}
V. S. Olkhovsky, E. Recami, and G. Salesi, Europhys. Lett. {\bf 57}, 879 (2002).
\bibitem{WinWin}
H. G. Winful, Nature {\bf 424}, 638 (2003).
\bibitem{Winful}
H. G. Winful, Phys. Rev. Lett. {\bf 91}, 260401 (2003).
\bibitem{Olk04}
V. S. Olkhovsky, E. Recami and J. Jakiel, Phys. Rep. {\bf 398}, 133 (2004).
\bibitem{Ber08}
A. E. Bernardini, Eur. Phys. J. {\bf C53}, 673 (2008).
\bibitem{Del03}
F. Delgado {\it et al.}, Phys. Rev. {\bf A68}, 032101 (2003).
\bibitem{Cal99}
A. Calogeracos and N. Dombey, Int. J. Mod. Phys. {\bf A14}, 631 (1999).
\bibitem{Dom99}
N. Dombey and A. Calogeracos, Phys. Rep. {\bf 315}, 41 (1999).
\bibitem{Che02}
Chun-Fang Li and Xi Chen, Ann. Phys. {\bf 12} (Leipzig), Ed.10, 916 (2002).
\bibitem{Ber07A}
A. E. Bernardini, to appear in J. Phys. {\bf A},
arXiv:0706.3930 [quant-ph].
\bibitem{Kle29}
O. Klein, Z. Phys. {\bf 53}, 157 (1929).
\bibitem{Zub80}
C. Itzykson and J. B. Zuber, {\it Quantum Field Theory} (Mc Graw-Hill Inc., New York, 1980).
\bibitem{Wig55}
E. P. Wigner, Phys. Rev. {\bf 98}, 145 (1955).
\bibitem{Hau89}
E. H. Hauge and J. A. Stovneng, Rev. Mod. Phys. {\bf 61}, 917 (1989).
\bibitem{Win03}
H. G. Winful, Phys. Rev. {\bf E68}, 016615 (2003).
\bibitem{Har62}
T. E. Hartman, J. Appl. Phys. {\bf 33}, 3427 (1962).
\bibitem{Aux1}
R. K. Su, G. Siu and X. Chou, J. Phys. {\bf A26}, 1001 (1993);
B. R. Holstein, Am. J. Phys. {\bf 66}, 507 (1998);
H. Nitta, T. Kudo and H. Minowa, Am. J. Phys. {\bf 67}, 966 (1999).
\bibitem{Kre04}
P. Krekora, Q. Su and R. Grobe, Phys. Rev. Lett. {\bf 92}, 040406 (2004);
\bibitem{Kre01}
P. Krekora, Q. Su  and R. Grobe, Phys. Rev. {\bf A63}, 032107 (2001).
\bibitem{Pet03}
V. Petrillo and D. Janner, Phys. Rev. {\bf A67}, 012110 (2003).
\bibitem{Lan89}
R. Landauer, Nature {\bf 341}, 567 (1989).
\bibitem{Ber06}
A. E. Bernardini, Phys. Rev. {\bf A74}, 062111 (2006).
\bibitem{Gav84}
B. Gaveau {\it et al.}, Phys. Rev. Lett. {\bf 53}, 419 (1984).
\bibitem{Smi60}
F. T. Smith, Phys. Rev. {\bf 118}, 349 (1960).
\bibitem{Kat06}
M. I. Katsnelson, K. S. Novoselov and A. K. Geim, Nature Phys. {\bf 02}, 620 (2006).
\bibitem{Gre85}
W. Greiner, B. Mueller, and J. Rafelski, {\em Quantum Electrodynamics of Strong Fields} (Springer, Berlin, 1985).
\bibitem{Pag05}
D. N. Page, New J. Phys. {\bf 7}, 203 (2005).
\end{thebibliography}
\end{document}